\documentclass[11pt,a4paper]{article}
\usepackage{jheppub}
\usepackage[utf8]{inputenc}
\usepackage[british]{babel}
\usepackage{lmodern}
\usepackage{dsfont}
\usepackage{amsmath}
\usepackage{amssymb}
\usepackage{amsfonts}
\usepackage{epsfig}
\usepackage{epstopdf}
\usepackage{graphicx,psfrag}
\usepackage[FIGBOTCAP,TABBOTCAP,tight]{subfigure}
\usepackage{enumitem}
\usepackage{array}
\usepackage{tikz}

\usetikzlibrary{arrows}
\usetikzlibrary{positioning}

\usepackage{multirow}
\usepackage{cellspace} %
\newcommand{\beq}{\begin{equation}}
\newcommand{\eeq}{\end{equation}}
\newcommand{\bea}{\begin{eqnarray}}
\newcommand{\eea}{\end{eqnarray}}

\newcommand{\eq}{equation~}
\newcommand{\eqs}{equations~}
\newcommand{\fig}{figure~}
\newcommand{\Fig}{Figure~}
\newcommand{\figs}{figures~}
\newcommand{\tbl}{table~}

\newcommand{\tr}{{\rm Tr}}

\def\lsi{\raise0.3ex\hbox{$<$\kern-0.75em\raise-1.1ex\hbox{$\sim$}}}
\def\gsi{\raise0.3ex\hbox{$>$\kern-0.75em\raise-1.1ex\hbox{$\sim$}}}

\newcommand*{\indeg}{n_{\mathrm{in}}}
\newcommand*{\outdeg}{n_{\mathrm{out}}}
\newcommand*{\multiind}[1]{\underline{#1}}

\graphicspath{{./figures/}{./data_su3paper/}}

\newcolumntype{M}[1]{>{\centering\arraybackslash}m{#1}}

\tikzset{vertex/.style={fill,draw,circle,inner sep=0pt,minimum size=3.5pt}}
\tikzset{ext-vertex/.style={draw,circle,inner sep=0pt,minimum size=3.5pt}}
\tikzset{edge/.style={->,> = latex'}}

\title{The SU(3) spin model with chemical potential by series expansion techniques}

\author[a]{Jangho Kim}
\author[a]{Anh Quang Pham}
\author[a,b]{Owe Philipsen}
\author[a,b]{Jonas Scheunert}
\affiliation[a]{Institut f\"ur Theoretische Physik,
Goethe-Universit\"at Frankfurt am Main, \\ Max-von-Laue-Str. 1, 60438 Frankfurt am Main, Germany}
\affiliation[b]{Helmholtz Research Academy Hesse for FAIR,\\ Max-von-Laue-Str. 12, 60438 Frankfurt am Main, Germany}

\emailAdd{jkim, pham, philipsen, scheunert@itp.uni-frankfurt.de}

\abstract{
The $SU(3)$ spin model with chemical potential  corresponds to 
a simplified version of QCD with static quarks in the strong coupling regime.
It has been studied previously as a 
testing ground for new methods aiming to overcome the sign problem of lattice QCD.
In this work we show that the equation of state and the phase structure of the 
model can be fully determined to reasonable accuracy by a linked cluster expansion. In particular, we compute the free energy to
14-th order in the nearest neighbour coupling. The resulting predictions for the equation of state and the location
of the critical end points agree with numerical determinations to ${\cal O}(1\%)$ and ${\cal O}(10\%)$, respectively.
While the accuracy for the critical couplings is still limited at the current series depth, the approach is
equally applicable at zero and non-zero imaginary or real chemical potential, as well as to effective QCD Hamiltonians obtained
by strong coupling and hopping expansions.
}

\begin{document}
\maketitle


\section{Introduction}

Despite growing demand from various fields of physics, the details of the QCD phase diagram as a function of
temperature $T$ and baryon chemical potential $\mu_B$ remain unknown to date. This is because of a severe sign problem 
due to the complex fermion determinant for $\mu_B\neq 0$, 
which prohibits straightforward Monte Carlo simulations of 
lattice QCD (for introductions, see, e.g., \cite{latbook,houches}). 
Controlled results only exist by indirect  
methods for the low density region $\mu_B\lsi 3T$, 
where no sign of criticality is observed \cite{lat18,lat19}.

This has motivated the search for alternative formulations and algorithms to solve this problem. While there is a vast literature on 
the general subject of sign problems, all approaches devised so far work 
for limited classes of Hamiltonians,
which do not (yet) include QCD. 
The purpose of the present paper is to test series expansion techniques, 
generically known as `high temperature' expansions in the condensed matter literature,
against the known numerical results for the $SU(3)$ spin model. 

The $SU(3)$ spin model with chemical potential is a system often used in the literature
to test new methods aiming at the QCD sign problem. 
It corresponds to QCD with a simplified determinant for static quarks
near the strong coupling limit. It features a local $SU(3)$ symmetry whose center 
$Z(3)$ spontaneously breaks at some critical coupling, as well as a sign problem at 
$\mu\neq 0$.
The model was reformulated free of a sign problem in terms of a flux 
representation \cite{gat1}, 
allowing for simulations \cite{gat2} by means of a worm algorithm. 
Likewise, simulations using a complex Langevin 
algorithm have been successful \cite{kw,bilic,aarts}. 
The phase structure of the $SU(3)$ spin model is therefore
known at zero and finite chemical potential with good precision.

Being analytic in nature, series expansion techniques are insensitive to sign problems and in principle
applicable to any Hamiltonian resembling a spin model. 
This is of particular relevance to finite density QCD, 
where combined strong coupling and hopping expansions produce effective Polyakov loop 
actions which closely resemble the model considered here \cite{llp,fromm,bind,Glesaaen:2015vtp}.
If such series expansions can be pushed to sufficiently high
orders for a required accuracy, the extension to $\mu_B\neq 0$ poses no 
fundamentally new problem.  In this work we demonstrate that for the 
$SU(3)$ spin model the equation of state as well as the phase diagram can be
determined to satisfactory accuracy with a reasonable effort. 

The paper is organised as follows. Section \ref{sec:model} introduces the spin model as well as the observables which we use
to characterise its phase diagram. In section \ref{sec:lce} we briefly summarise 
the main features of the linked cluster expansion and
apply it to the spin model at hand. From the resulting power series, 
we then extract the equation of state and compare with Monte Carlo results at zero density. 
We discuss how to extract 
the phase structure and its critical features by means of series analyses
in section \ref{sec:pda}.
Finally, section \ref{sec:conc}
concludes with a discussion of the prospects of this approach for effective theories of QCD.

\section{The \texorpdfstring{$SU(3)$}{SU(3)} spin model\label{sec:model}}

\begin{figure}[t]
 \centerline{
 \includegraphics[width=0.6\textwidth]{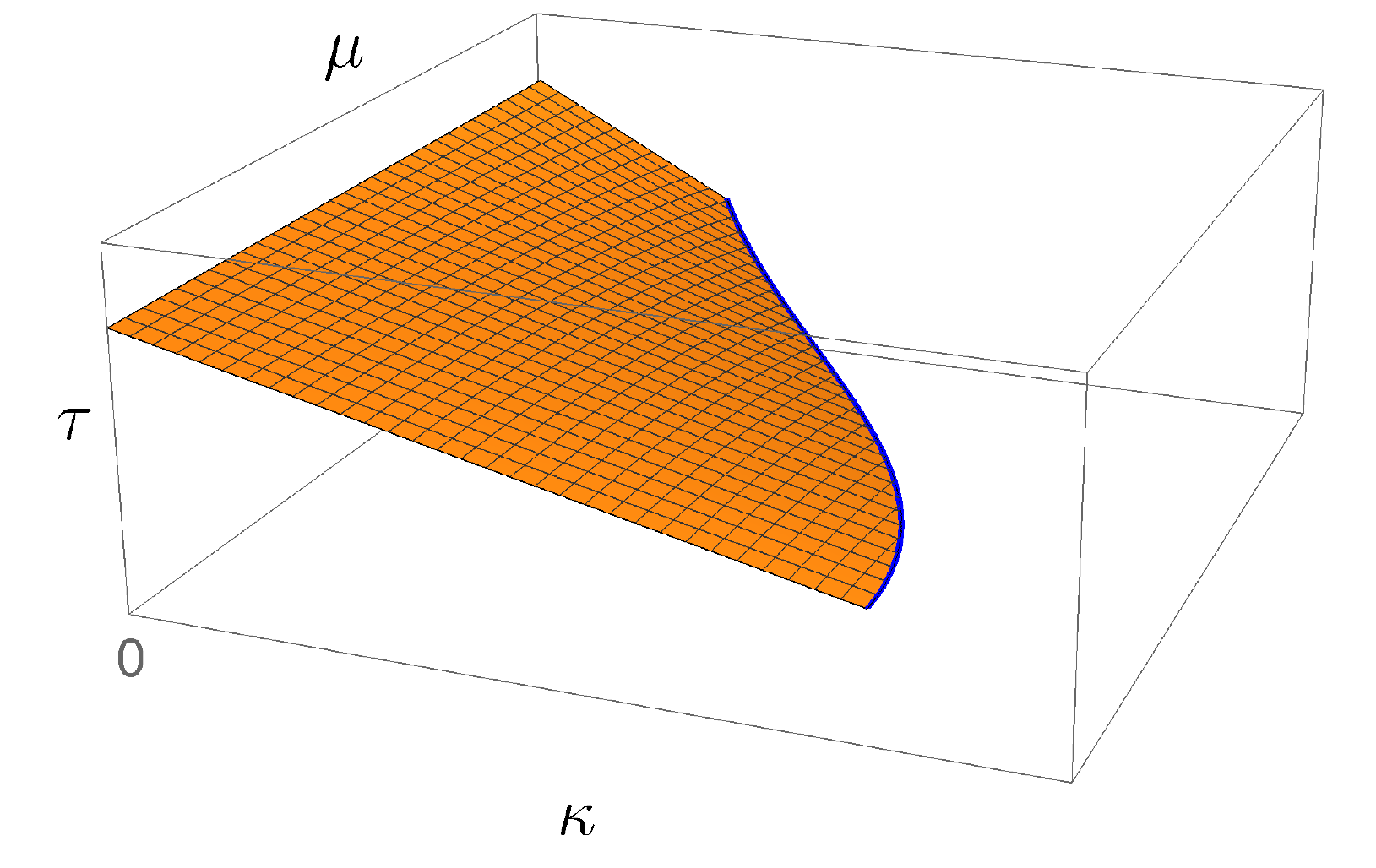}
  }
  \caption{Schematic phase diagram of the $SU(3)$ spin model. A surface of first-order transitions is bounded by a $Z(2)$-critical
  line beyond which the transitions are analytic crossover.}
  \label{fig:schem}
\end{figure}

We consider the action
\beq\label{eq:su3spinaction}
S=-\sum_x \Big(\sum_{k=1}^3\tau \left[L(x)L^*(x+\hat{k})+L^*(x)L(x+\hat{k})\right]+\kappa \left[e^\mu L(x)+e^{-\mu}L^*(x)\right]\Big) \;,
\eeq
with $x$ denoting the sites of a three-dimensional cubic lattice with  
$\hat{k}$ the unit vectors in the three directions. 
The fields $L(x)=\tr W(x)$ are complex scalar variables, resulting 
by taking the trace of the $SU(3)$-matrices $W(x)$. 
The analogy to other spin models is emphasised by defining 
 \beq
 \eta=\kappa e^\mu,\; \quad \bar{\eta}=\kappa e^{-\mu}\;.
 \eeq
 Then $\langle L\rangle$ plays the role of magnetisation and
$\eta,\bar{\eta}$ represent symmetry breaking external fields.
For $\eta=\bar{\eta}=0$ the action is invariant under global transformations  
\beq
L'=zL,\quad z\in Z(3)=\{1,e^{i\frac{\pi}{3}},e^{i\frac{2\pi}{3}}\}\;,
\eeq
with $Z(3)$ the center of $SU(3)$.
When the model is viewed as
an effective theory for lattice QCD, the $W(x)$ are temporal Wilson lines representing 
the colour propagation of static quarks,
the nearest neighbour coupling $\tau$ is related to temperature and the gauge coupling, 
while $\kappa$ parametrises the quark mass. 
For our analytic calculations, the lattice can be considered infinitely large.
The corresponding grand-canonical partition function is given by the functional integral
\beq
Z(\tau,\kappa,\mu)=\int DW \; e^{-S[W]}\;.
\eeq
It is obvious that for $\mu=0$ the action is real. The partition function then is 
straightforward to simulate by standard Metropolis methods, which 
we will use as a benchmark for our series expansions. In this case we work with 
finite volumes and periodic boundary conditions
in all directions. For $\mu\neq 0$, the action is complex and the simulations exhibit 
a sign problem. 
The phase structure can be inferred from previous work curing the sign problem \cite{kw,bilic,aarts,gat2} and is shown
schematically in figure \ref{fig:schem}. Note that, when the continuous $SU(3)$ variables $L(x)$ are restricted to their center
elements in $Z(3)$, the model reduces to the discrete 3-state Potts model in three dimensions, which has been studied both numerically
\cite{Alford:2001ug,Mercado:2011ua} as well as by expansion methods \cite{Guttmann:1993ad,Hellmund:2003vw,Hellmund:2006yz}.
Its phase diagram looks qualitatively the same, here we focus on the QCD-like $SU(3)$ situation. 

The natural observable to compute by series expansion is the free energy density
\beq
  f = - \frac{\log(Z)}{V}\;,
 \eeq
 from which all other thermodynamic quantities can be derived. For comparison with Monte Carlo simulations,
 which cannot evaluate the free energy directly,  we study the combination of  first derivatives
 \beq
  \Delta S  = - \frac{\partial f}{\partial \tau} - \frac{\partial f}{\partial \eta}\;, 
  \label{eq:int}
 \eeq 
 which is related to the trace of the energy momentum tensor and in QCD is 
 called interaction measure.
 In order to identify phase transitions, we look for maximal fluctuations in the spin model analogues of the magnetic susceptibility and the 
 specific heat, respectively, 
 \bea
  \chi &=&
  - \frac{\partial^2 f}{\partial \eta^2}
  - \frac{\partial^2 f}{\partial \bar{\eta}^2}
  - 2 \frac{\partial^2 f}{\partial \eta \partial \bar{\eta}}\;,\label{eq:susc}\\
  C&=&-\frac{\partial^2 f}{\partial\tau^2}\;.
  \label{eq:specheat}
\eea

\section{Linked cluster expansion \label{sec:lce}}

There are many expansions in physics that can be expressed in terms of linked
clusters (connected graphs). For an introduction to the general subject 
see \cite{Oitma:2006aa}, where 
the method we are employing is also called the `free graph expansion', 
a name which will become clear later.
One famous application of this expansion is to the $\phi^4$ theory \cite{Baker:1981zz},
where it
was used to prove triviality \cite{Luscher:1988gk}. Extensions
to very high orders in this case can be found in \cite{Reisz:1995ag}. 
Following these references, we employ a so-called
`vertex-renormalisation'\footnote{Not to be confused with the renormalisation
in quantum field theories to remove ultra-violet divergences}, where
the number of graphs one has to consider is decreased at the cost of increased
algebraic complexity. Further improvements can be achieved by an 
additional `bond-renormalisation'. An impressive
example of applying both renormalisation techniques is
\cite{Campostrini:2000yf}, where two-point Green's functions for a generalised
3D Ising model are obtained to 25th order on the {\it bcc} lattice and to 23rd order
on the {\it sc} lattice.
Our application follows closely the review article by Wortis \cite{Wortis:1980zc}, which
gives a clear explanation of the method.

\subsection{Graphs}

In this section we review what is necessary to give a precise meaning to the
involved formulas. Our notation and definitions are based on a combination of
those introduced in
\cite{Luscher:1988gk,Campostrini:2000yf,ESSAM:1970zz,Epp:2010aa}. One
complication we encounter is that, due to the particular form of the
interaction term in the spin model, we have to consider directed graphs. A
directed graph consists of two finite sets, a set $V(G)$ of vertices and a set
$D(G)$ of directed bonds together with two mappings
\begin{align}
  i &: D(G) \rightarrow V(G),\\
  t &: D(G) \rightarrow V(G).
\end{align}
In this way, each bond $b\in D(G)$ is associated with the ordered pair
$(i(b),t(b))$ of endpoints, where $i(b)$ is called the initial and $t(b)$ is
called the terminal point of $b$. The bond can be visualised as an arrow with
tail $i(b)$ and head $t(b)$. Two vertices, which are the endpoints of
the same bond, are adjacent to each other.

The in-degree $\indeg(v)$ of a vertex is the number of directed bonds with
$t(b)=v$ and the out-degree $\outdeg(v)$ is defined analogously. Their sum
gives the degree of a vertex, $n(v)=\indeg(v)+\outdeg(v)$.

Sometimes one distinguishes external/rooted vertices $R(G)$ and internal
vertices $I(G)$, $V(G)=R(G)\cup I(G)$. An $r$-rooted graph denotes a graph with
$r$ external vertices and we will always assume that the external vertices have
the labels $1,\ldots,r$.

On the level of graphs, the external and internal vertices differ in how they
enter the definition of a graph isomorphism. Two graphs $G$ and $G'$ are
considered isomorphic if there exist two mappings
\begin{align}
  \varphi &: V(G) \rightarrow V(G'), \\
  \lambda &: D(G) \rightarrow D(G'),
\end{align}
with the properties that
\begin{enumerate}[label=(\alph*)]
  \item $\varphi$ and $\lambda$ are bijective,
  \item $\varphi(r) = r$ for all $r\in R(G)$,
  \item $i\circ\lambda(b)=\varphi\circ i(b) \land 
    t\circ\lambda(b)=\varphi\circ t(b)$ for all $b\in D(G)$.
\end{enumerate}
From now on, whenever we define a set $S$ of graphs, we always actually mean the
quotient set $S/\sim$, where $G\sim G'$ if they are isomorphic.
The symmetry factor $S(G)$ of a graph $G$ is the order of the automorphism
group of $G$.

Two vertices $v_1$ and $v_2$ are called connected, if there is a sequence of
adjacent vertices $w_1,\ldots,w_n$, with $w_1=v_1$ and $w_n=v_2$. A graph is
defined to be connected if its vertices are pairwise connected.

We write $G\setminus v$ for the deletion of the vertex $v$ from the graph $G$.
This means that $G\setminus v$ is obtained from $G$ by removing $v$ from $V(G)$
and removing all bonds $b$ from $E(G)$ with $t(b)=v$ or $i(b)=v$. For a
connected graph with external vertices, an articulation point is a vertex $v$
with the property that $G\setminus v$ has a vertex which is not connected to a
root. In the case of a graph with no external vertices, an articulation point
is a vertex $v$ such that $G\setminus v$ is disconnected. A 1-irreducible graph
is then a graph with no articulation points.

Finally, a 1-insertion denotes a connected graph with one external vertex,
where deleting the external vertex leaves the graph connected. In case of a
1-insertion we define 1-irreducibility to mean that there is no articulation
point except of the root (otherwise, the 1-irreducible 1-insertions would only
consist of one single vertex graph).

\subsection{Unrenormalised expansion}

We use the linked cluster expansion to determine the Taylor expansion of the
free energy of the spin model in $\tau$ around $\tau=0$ in terms of connected
graphs. The general graphs which enter this expansion are valid for many
different theories, so one has to make a connection between the graphs and the
concrete theory. This is achieved by assigning weights $W^0(G)$ to the graphs.
These weights are expressed in so-called bare semi-invariants, which are the
cumulant correlations of the theory without interactions.

In the case of the $SU(3)$ spin model the bare semi-invariants
$M_{n,m}^0(\eta,\bar{\eta})$ with $n,m\in \mathds{N}$, $\eta=\kappa
\exp(\mu)$ and $\bar{\eta}=\kappa \exp(-\mu)$ can be obtained via
\begin{equation}
  M_{n,m}^0(\eta,\bar{\eta}) =
  \left(\frac{\partial}{\partial\eta}\right)^n
  \left(\frac{\partial}{\partial\bar{\eta}}\right)^m
  \log\left[\int_{SU(3)} dW \, \exp(\eta L + \bar{\eta} L^\ast)\right].
\end{equation}
It is useful to have an expansion of the logarithm in terms of $\eta$ and $\bar{\eta}$.
To this end, consider
\begin{equation}
  z(\alpha,\eta,\bar{\eta}) =
  \int_{SU(3)} dW \, \exp[\alpha(\eta L + \bar{\eta} L^\ast)].
\end{equation}
Using Fa\`{a} di Bruno's formula one then obtains
\begin{equation}
  \log[z(\alpha,\eta,\bar{\eta})] =
  \sum_{n=1}^{\infty} \sum_{k=1}^{n} (-1)^{k+1} \frac{(k-1)!}{n!}
  B_{n,k}\left(\left.\frac{\partial z}{\partial a}\right|_{\alpha=0}, \ldots,
      \left.\frac{\partial^{n-k+1}z}{\partial\alpha^{n-k+1}}\right|_{\alpha=0}\right)
  \alpha^n,
\end{equation}
where $B_{n,k}$ denotes the partial Bell polynomial
\begin{equation}
  B_{n,k} = 
  \sum_{j\in\mathds{N}^{n-k+1}}
  \left[\sum_{i=1}^{n-k+1} j_i = k\right]
  \left[\sum_{i=1}^{n-k+1} i j_i = n\right]
  n!
  \prod_{i=1}^{n-k+1} \frac{1}{j_i} \left(\frac{x_i}{i!}\right)^{j_i}.
\end{equation}
We use the notation that any logical statement in square brackets evaluates to
1 if it is true and 0 otherwise. The derivatives of $z$ can be computed via
\begin{equation}
  \left. \frac{\partial^p z}{\partial \alpha^p} \right|_{\alpha=0} =
  \sum_{q=0}^p \binom{p}{q} I_{q,p-q} \eta^q \bar{\eta}^{p-q}
\end{equation}
with the $SU(3)$-integral
\begin{equation}
  I_{q,p-q} = \int_{SU(3)} dW L^q L^{\ast (p-q)},
\end{equation}
for which an explicit formula is given in \cite{gat1}.

For a directed graph $G$ we can now define its weight $W^0(G)$ to be
\begin{equation}
  W^0(G) = \prod_{v\in I(G)} M_{\outdeg(v),\indeg(v)}^0.
\end{equation}

So far, all definitions are independent of the underlying lattice of the
theory. In the linked cluster expansion, the information about the lattice is
encoded in the free embedding numbers, also called free multiplicities. On a
three-dimensional cubic lattice, the free embedding number counts the number of
ways to put a graph on $\mathds{Z}^3$, with one vertex fixed to the origin
while for all others adjacent vertices have to correspond to nearest neighbours
on the lattice. For the free embedding number it is allowed to place two
vertices on the same lattice point, which makes its computation relatively easy
in comparison to other embedding numbers.
To give a more formal definition, let $X_{G,v}$ denote the set of all functions
that map all vertices of the graph $G$ to $\mathds{Z}^3$ with $x(v)=0$ for all
$x\in X_{G,v}$. Then the free embedding number $M(G)$ of a graph $G$ is defined
to be (with $v\in V(G)$ chosen arbitrarily)
\begin{equation}
  M(G) = \sum_{x\in X_{G,v}} \prod_{b\in D(G)} 
  \Big[  d(x \circ t(b), x \circ i(b)) = 1 \Big],
\end{equation}
where $d$ denotes the lattice distance
\begin{equation}
  d(x,y) = \sum_{\mu=1}^3 \lvert x_\mu - y_\mu \rvert.
\end{equation}

We are now able to write the expansion of the free energy density in terms of
connected graphs. Define $\mathcal{G}_c^{(i)}$ to be the set of all connected and
directed graphs with $i$ bonds and no external vertices, then the free energy
density evaluates to
\begin{equation}
  -f = \lim_{V\rightarrow\infty} \frac{\log(Z)}{V} =
  \sum_{i=0}^{\infty} \tau^i \sum_{G\in\mathcal{G}_c^{(i)}}
  \frac{M(G)W^0(G)}{S(G)}.
\end{equation}
To be a bit more explicit, we evaluate this formula to $\mathcal{O}(\tau^2)$ as an example.
The involved graphs, their symmetry and embedding numbers and their weights are
listed in \tbl \ref{tab:free-energy-graphs-unrenormalised}. Consequently, one has
\begin{align}
  \label{eq:free-energy-truncated}
  \begin{split}
    -f 
    & =
    M^0_{0,0} +
    \tau \left(
      6 M^0_{1,0} M^0_{0,1}
    \right) + \\
    & \qquad
    \tau^2 \biggl(
      3 M^0_{2,0} M^0_{0,2} + 3 (M^0_{1,1})^2 +
      36 M^0_{1,0} M^0_{1,1} M^0_{0,1} + \\
      & \qquad \qquad
      18 (M^0_{1,0})^2 M^0_{0,2} +
      18 (M^0_{0,1})^2 M^0_{2,0}
    \biggr).
  \end{split}
\end{align}
The bare semi-invariants can then be evaluated according to the explanation
at the beginning of this section.

\begin{table}[t]
  \centering
  \begin{tabular}{M{2.5cm} M{2.5cm} M{2.5cm} M{2.5cm}}
    Graph & $S$ & $M$ & $W^0$ \\
    \hline
    \begin{tikzpicture}
      \node[vertex] (a) at (0,0) {};
    \end{tikzpicture} &
    $1$ & $1$ & $M^0_{0,0}$ \\
    \begin{tikzpicture}
      \node[vertex] (a) at (0,0) {};
      \node[vertex] (b) at (0,0.5) {};
      \draw[edge] (a) to (b);
    \end{tikzpicture} &
    $1$ & $2d$ & $M^0_{1,0} M^0_{0,1}$ \\
    \begin{tikzpicture}
      \node[vertex] (a) at (0,0) {};
      \node[vertex] (b) at (0,0.5) {};
      \draw[edge] (a) to[bend left] (b);
      \draw[edge] (a) to[bend right] (b);
    \end{tikzpicture} &
    $2$ & $2d$ & $M^0_{2,0} M^0_{0,2}$ \\
    \begin{tikzpicture}
      \node[vertex] (a) at (0,0) {};
      \node[vertex] (b) at (0,0.5) {};
      \draw[edge] (a) to[bend left] (b);
      \draw[edge] (b) to[bend left] (a);
    \end{tikzpicture} &
    $2$ & $2d$ & $(M^0_{1,1})^2$ \\
    \begin{tikzpicture}
      \node[vertex] (a) at (0,0) {};
      \node[vertex] (b) at (0.25,0.5) {};
      \node[vertex] (c) at (0.5,0) {};
      \draw[edge] (a) to (b);
      \draw[edge] (b) to (c);
    \end{tikzpicture} &
    $1$ & $(2d)^2$ & $M^0_{1,0} M^0_{1,1} M^0_{0,1}$ \\
    \begin{tikzpicture}
      \node[vertex] (a) at (0,0) {};
      \node[vertex] (b) at (0.25,0.5) {};
      \node[vertex] (c) at (0.5,0) {};
      \draw[edge] (a) to (b);
      \draw[edge] (c) to (b);
    \end{tikzpicture} &
    $2$ & $(2d)^2$ & $(M^0_{1,0})^2 M^0_{0,2}$ \\
    \begin{tikzpicture}
      \node[vertex] (a) at (0,0) {};
      \node[vertex] (b) at (0.25,0.5) {};
      \node[vertex] (c) at (0.5,0) {};
      \draw[edge] (b) to (a);
      \draw[edge] (b) to (c);
    \end{tikzpicture} &
    $2$ & $(2d)^2$ & $(M^0_{0,1})^2 M^0_{2,0}$ \\
  \end{tabular}
  \caption{Graphs considered for the free energy up to $\mathcal{O}(\tau^2)$
    with their symmetry numbers $S$, embedding numbers $M$ on a $d$-dimensional
  square lattice and weights.}
  \label{tab:free-energy-graphs-unrenormalised}
\end{table}

\subsection{Vertex-renormalised expansion \label{subsec:vertex-renorm}}

The set $\mathcal{G}_c^{(i)}$ becomes large rather fast when $i$ is increased.
As mentioned above, here we employ a method called vertex-renormalisation,
which reduces the number of graphs one has to consider.

To this end, we first define the self fields, which collect all possible ways to
decorate a vertex (note that the free embedding number factorises along
articulation points). Denoting by $\mathcal{I}_{(n,m)}^{(i)}$ all
one-insertions with $i$ edges and where the external vertex has out-degree $n$
and in-degree $m$, the self field $G_{n,m}$ is defined to be (remember that
$W^0$ was defined in a way such that external vertices give no contribution)
\begin{equation}
  \label{eq:self-fields-unrenormalised}
  G_{n,m} =
  \sum_{i=n+m}^{\infty}
  \tau^i \sum_{G\in\mathcal{I}_{(n,m)}^{(i)}} \frac{M(G) W^0(G)}{S(G)}.
\end{equation}
This in turn can be used to establish the notion of renormalised
semi-invariants
\begin{align}
  \label{eq:renormalised-semi-invariants}
  \begin{split}
    M_{n,m}
    & =
    M_{n,m}^0 +
    \sum_{p=1}^\infty \frac{1}{p!} 
    \sum_{(l_1,k_1),\ldots,(l_k,k_p) \in \mathds{N}^2\setminus(0,0)}
    \left(\prod_{j=1}^{p} G_{l_j,k_j}\right)
    M_{n+l_1+\ldots +l_p, m+k_1+\ldots +k_p}^{0} \\
    & =
    \exp\left(\sum_{(l,k) \in \mathds{N}^2\setminus(0,0)}
      G_{l,k}\left(\frac{\partial}{\partial \eta}\right)^{l}
    \left(\frac{\partial}{\partial \bar{\eta}}\right)^{k}\right)
    M_{n,m}^{0}.
  \end{split}
\end{align}
Defining the renormalised weight $W(G)$ of a graph in the obvious way,
one can obtain the free energy density via
\begin{equation}
  \label{eq:free-energy-renormalised}
  -f =
  M_{0,0} + \Phi - \sum_{l,k \in \mathds{N}^+} G_{l,k} M_{l,k}
\end{equation}
where the so called $\Phi$-functional
\begin{equation}
  \label{eq:phi-functional}
  \Phi = 
  \sum_{i=1}^{\infty} \tau^i \sum_{G\in\mathcal{G}_{\mathrm{1}}^{(i)}}
  \frac{M(G) W(G)}{S(G)}
\end{equation}
contains sums over $\mathcal{G}_{\mathrm{1}}^{(i)}$, the sets of all
1-irreducible graphs with no external vertices and $i$ edges. Those sets are
much smaller than $\mathcal{G}_c^{(i)}$.

At this point, however, nothing is gained yet, because in order to obtain the
self fields one still has to evaluate sums over all 1-insertions (also those
which are not 1-irreducible) in \eq \eqref{eq:self-fields-unrenormalised}. 
Replacing all bare semi-invariants by their renormalised counter parts in this
equation enables one to restrict the sums to 1-irreducible 1-insertions
$\mathcal{I}_{1,(n,m)}^{(i)}$
\begin{equation}
  \label{eq:self-fields-renormalised}
  G_{n,m} =
  \sum_{i=n+m}^{\infty} \tau^i 
  \sum_{G\in\mathcal{I}_{1,(n,m)}^{(i)}} \frac{M(G) W(G)}{S(G)}.
\end{equation}
As a result, however, \eqs \eqref{eq:renormalised-semi-invariants} and
\eqref{eq:self-fields-renormalised} are coupled equations. Therefore, the
following iterative strategy is used: suppose one wants to obtain the free
energy to the order $\sim\tau^{n_{\max}}$. Using \eq \eqref{eq:free-energy-renormalised}
this necessitates the determination of the renormalised semi-invariants
$M_{n,m}$ to order $n_{\max}-n-m$. To leading order
\begin{equation}
  \label{eq:renormalised-semi-invariants-leading-order}
  M_{n,m} = M_{n,m}^0 + \mathcal{O}(\tau),
\end{equation}
which enables the determination of $G_{1,0}$ and $G_{0,1}$ to order $1$ in
$\tau$ using \eq \eqref{eq:self-fields-renormalised}. These self fields can
then be used to determine the renormalised semi-invariants to order 1, and in
this way the procedure can be continued to the necessary order. In general,
having obtained all renormalised semi-invariants $M_{n,m}$ to order $p$ with
$(n+m)\leq p$, one can determine the $G_{n,m}$ to order $p+1$ with $(n+m)\leq
p+1$ and use those to obtain the renormalised semi-invariants to order $p+1$.

We again illustrate the procedure by deriving the free energy to order
$\mathcal{O}(\tau^2)$. At first, we have to determine the
renormalised semi-invariants $M_{n,m}$ and the self-fields $G_{n,m}$ to
$\mathcal{O}(\tau^2)$.
Since the leading contribution of $G_{n,m}$ is of $\mathcal{O}(\tau^{n+m})$ the
relevant contributions from \eq \eqref{eq:renormalised-semi-invariants} 
to the renormalised semi-invariants are
\begin{align}
  \label{eq:renormalised-semi-invariants-truncated}
  \begin{split}
    M_{n,m} =
    & M_{n,m}^0 + \\
    & G_{0,1} M_{n,m+1}^0 + G_{1,0} M_{n+1,m}^0 + \\
    & G_{0,2} M_{n,m+2}^0 + G_{2,0} M_{n+2,m}^0 + \\
    & (G_{1,1} + G_{0,1} G_{1,0}) M_{n+1,m+1}^0 + \\
    & \frac{1}{2} G_{0,1}^2 M_{n,m+2}^0 + \frac{1}{2} G_{1,0}^2 M_{n+2,m}^0.
  \end{split}
\end{align}
Furthermore, the relevant 1-irreducible 1-insertions for the self-fields
are shown in \tbl \ref{tab:self-field-graphs-renormalised} and therefore
\begin{align}
  \label{eq:self-fields-renormalised-truncated-first}
  G_{1,0} & = 6 \tau M_{0,1} \quad \text{(exact!)}, \\
  G_{1,1} & = 6 \tau^2 M_{1,1} + \mathcal{O}(\tau^3), \\
  \label{eq:self-fields-renormalised-truncated-last}
  G_{2,0} & = 3 \tau^2 M_{0,2} + \mathcal{O}(\tau^3).
\end{align}
Note that we give the self fields $G_{n,m}$ only for $n \geq m$, because
one has the symmetries $M_{n,m}(\eta,\bar{\eta}) = M_{m,n}(\bar{\eta},\eta)$
and $G_{n,m}(\eta,\bar{\eta}) = G_{m,n}(\bar{\eta},\eta)$.
To decouple \eq \eqref{eq:renormalised-semi-invariants-truncated} from \eqs
\eqref{eq:self-fields-renormalised-truncated-first} to
\eqref{eq:self-fields-renormalised-truncated-last} we use the leading
order expression for the renormalised semi-invariants \eq
\eqref{eq:renormalised-semi-invariants-leading-order} and get from
\eq \eqref{eq:self-fields-renormalised-truncated-first}
\begin{equation}
  G_{1,0} = 6 \tau M_{0,1}^0 + \mathcal{O}(\tau^2).
\end{equation}
This in turn can be used in \eq
\eqref{eq:renormalised-semi-invariants-truncated} to obtain
\begin{align}
  M_{0,0} & =
  M_{0,0}^0 + 12 \tau M_{0,1}^0 M_{1,0}^0 + \mathcal{O}(\tau^2), \\
  M_{1,0} & =
  M_{1,0}^0 +
  6 \tau M_{0,1}^0 M_{2,0}^0 + 
  6 \tau M_{1,0}^0 M_{1,1}^0 + \mathcal{O}(\tau^2).
\end{align}
Inserting these equations into \eqs
\eqref{eq:self-fields-renormalised-truncated-first} to
\eqref{eq:self-fields-renormalised-truncated-last} results in
\begin{align}
  G_{1,0} & = 
  6 \tau M_{0,1}^0 +
  36 \tau^2 M_{0,1}^0 M_{1,1}^0 +
  36 \tau^2 M_{1,0}^0 M_{0,2}^0 
  + \mathcal{O}(\tau^3), \\
  G_{1,1} & =
  6 \tau^2 M_{1,1}^0 + \mathcal{O}(\tau^3), \\
  G_{2,0} & =
  3 \tau^2 M_{0,2}^0 + \mathcal{O}(\tau^3),
\end{align}
which, finally, can be used together with \eq
\eqref{eq:renormalised-semi-invariants-truncated} to determine
\begin{equation}
  \begin{split}
    M_{0,0} =
    & M_{0,0}^0 + 12 \tau M_{0,1}^0 M_{1,0}^0 +\\
    & 6 \tau^2
    \left(M_{0,2}^0 M_{2,0}^0 + (M_{1,1}^0)^2 + 
      9 M_{0,2}^0 (M_{1,0}^0)^2 + (M_{1,0}^0)^2 M_{2,0}^0 + 
    M_{0,1}^0 M_{1,0}^0 M_{1,1}^0 \right).
  \end{split}
\end{equation}

Next one determines the $\Phi$-functional from \eq \eqref{eq:phi-functional}.
The relevant graphs are graphs $2$ to $4$ in \tbl
\ref{tab:free-energy-graphs-unrenormalised}, however with the bare semi-invariants
replaced by their renormalised counterparts for the weights:
\begin{equation}
  \Phi = 6 \tau M_{0,1} M_{1,0} + 
  3 \tau^2 \left( M_{0,2} M_{2,0} + M_{1,1}^2 \right).
\end{equation}
Inserting the expressions for the renormalised semi-invariants and the
self-fields in terms of the bare semi-invariants which were derived above
into \eq \eqref{eq:free-energy-renormalised} then results in the same
expression that was obtained in \eq \eqref{eq:free-energy-truncated}.

While at this point it might seem that the renormalised procedure introduces
unnecessary complications, we remark that for higher orders a significant reduction
of graphs is achieved.

\begin{table}[t]
  \centering
  \begin{tabular}{M{2.5cm} M{2.5cm} M{2.5cm} M{2.5cm}}
    Graph & $S$ & $M$ & $W$ \\
    \hline
    \vspace{2pt}
    \begin{tikzpicture}
      \node[ext-vertex] (a) at (0,0) {};
      \node[vertex] (b) at (0,0.5) {};
      \draw[edge] (a) to (b);
    \end{tikzpicture} &
    $1$ & $2d$ & $M_{0,1}$ \\
    \begin{tikzpicture}
      \node[ext-vertex] (a) at (0,0) {};
      \node[vertex] (b) at (0,0.5) {};
      \draw[edge] (b) to (a);
    \end{tikzpicture} &
    $1$ & $2d$ & $M_{1,0}$ \\
    \begin{tikzpicture}
      \node[ext-vertex] (a) at (0,0) {};
      \node[vertex] (b) at (0,0.5) {};
      \draw[edge] (a) to[bend left] (b);
      \draw[edge] (a) to[bend right] (b);
    \end{tikzpicture} &
    $2$ & $2d$ & $M_{0,2}$ \\
    \begin{tikzpicture}
      \node[ext-vertex] (a) at (0,0) {};
      \node[vertex] (b) at (0,0.5) {};
      \draw[edge] (b) to[bend left] (a);
      \draw[edge] (b) to[bend right] (a);
    \end{tikzpicture} &
    $2$ & $2d$ & $M_{2,0}$ \\
    \begin{tikzpicture}
      \node[ext-vertex] (a) at (0,0) {};
      \node[vertex] (b) at (0,0.5) {};
      \draw[edge] (a) to[bend left] (b);
      \draw[edge] (b) to[bend left] (a);
    \end{tikzpicture} &
    $1$ & $2d$ & $M_{1,1}$ \\
  \end{tabular}
  \caption{Graphs considered for the self fields up to $\mathcal{O}(\tau^2)$
    in the vertex-renormalised scheme. External vertices are depicted as
  unfilled circles.}
  \label{tab:self-field-graphs-renormalised}
\end{table}

\subsection{Notes about implementation}

The sum over the $l_i$ and $k_i$ in \eq \eqref{eq:renormalised-semi-invariants}
contains several terms which are equal. For an efficient evaluation of the sum,
it should be written in such way that it does not contain redundancies.
To this end, we introduce the following order-relation on multi-indices of
$\mathds{N}^k$: for $\multiind{n},\multiind{m} \in \mathds{N}^k$ one has
\begin{align}
  \multiind{n} = \multiind{m} 
  \Leftrightarrow &
  (n_1 = m_1, \ldots, n_k = m_k), \\
  \begin{split}
    \multiind{n} > \multiind{m}
    \Leftrightarrow &
    (\left|\multiind{n}\right| > \left|\multiind{m}\right|) \\
    & \lor
    \left(\left|\multiind{n}\right|=\left|\multiind{m}\right| \land 
      U = \lbrace i \vert n_i \neq m_i\rbrace \neq \emptyset \land
      n_i>m_i
    \text{ for } i = \min(U)\right).
  \end{split}
\end{align}
Then, for $\multiind{n}\in\mathds{N}^2$ one can rewrite
\eq \eqref{eq:renormalised-semi-invariants} as
\begin{equation}
  M_{\multiind{n}} =
  M_{\multiind{n}}^0 +
  \sum_{p=1}^\infty \frac{1}{p!} 
  \sum_{\multiind{l}_1 \geq \ldots \geq \multiind{l}_p > (0,0)}
  m(\multiind{l}_1, \ldots, \multiind{l}_p)
  \left(\prod_{j=1}^{p} G_{\multiind{l}_j}\right)
  M_{\multiind{n}+\multiind{l}_1+\ldots +\multiind{l}_p}^{0},
\end{equation}
where the combinatorial factor $m(\multiind{l}_1, \ldots, \multiind{l}_p)$
counts the number of unique permutations of its arguments.

For graph isomorphism checking and symmetry numbers we used the graph
procedures provided by McKay's \texttt{nauty}
\cite{McKay:2014aa}. The embedding numbers were computed in
\texttt{Mathematica} using the algorithm described in \cite{Luscher:1988gk}.
The computation of the graph weights and the vertex-renormalisation were also
done in \texttt{Mathematica}, among other things because it implements multi
precision arithmetic via the \texttt{GMP} \cite{Grandlund:2015aa} in a
convenient way. 
The entire sequence of steps of the calculation and their respective run times
are listed in appendix \ref{sec:a}.

\subsection{The equation of state}

\begin{table}[t]
  \centering
  \begin{tabular}{ l l l }
    $n$ & $a_n(\kappa=0.016, \mu=0)$ & $a_n(\kappa = 0.005, \mu = 1.53)$ \\
    \hline
    \vspace{2pt}
    $0$  & $2.5736524 \times 10^{-4}$ & $2.7052164 \times 10^{-5}$ \\
    $1$  & $1.5606716 \times 10^{-3}$ & $1.8693971 \times 10^{-4}$ \\
    $2$  & $3.0102733$                & $3.0014059$ \\
    $3$  & $1.0669117$                & $1.0105599$ \\
    $4$  & $2.2956382 \times 10^1$    & $2.2581260 \times 10^1$ \\
    $5$  & $3.6334683 \times 10^1$    & $3.3868779 \times 10^1$ \\
    $6$  & $3.5390482 \times 10^2$    & $3.3727542 \times 10^2$ \\
    $7$  & $1.1201151 \times 10^3$    & $1.0086932 \times 10^3$ \\
    $8$  & $8.0957880 \times 10^3$    & $7.3384124 \times 10^3$ \\
    $9$  & $3.6577000 \times 10^4$    & $3.1440809 \times 10^4$ \\
    $10$ & $2.3886115 \times 10^5$    & $2.0373416 \times 10^5$ \\
    $11$ & $1.2968389 \times 10^6$    & $1.0566197 \times 10^6$ \\
    $12$ & $8.2969242 \times 10^6$    & $6.6453550 \times 10^6$ \\
    $13$ & $4.9682475 \times 10^7$    & $3.8315611 \times 10^7$ \\
    $14$ & $3.2081191 \times 10^8$    & $2.4228700 \times 10^8$
  \end{tabular}
  \caption{Coefficients of the free energy \eqref{eq:fseries} for 
    specific values of $\kappa$ and $\mu$. The $a_n$ have been determined through
    $\mathcal{O}(\kappa^{60})$.}
  \label{tab:coefficients}
\end{table}
Using these tools, we derived the free energy through $\mathcal{O}(\tau^{14})$ and,
in order to keep the size of the expressions under control, all products of bare
semi-invariants were expanded and truncated at  $\mathcal{O}(\kappa^{60})$.
While the corresponding expressions are too long to print here,
for purposes of reproducibility we give the numerical values of the coefficients in the $\tau$-series,
\beq
 -f (\tau,\kappa,\mu)= \sum_{n=0}^{14} a_n(\kappa,\mu) \tau^n\;,
 \label{eq:fseries}
\eeq
evaluated for two different choices of $(\kappa,\mu)$ in \tbl \ref{tab:coefficients}.

\begin{figure}[t]
 \centerline{
 \includegraphics[width=0.5\textwidth]{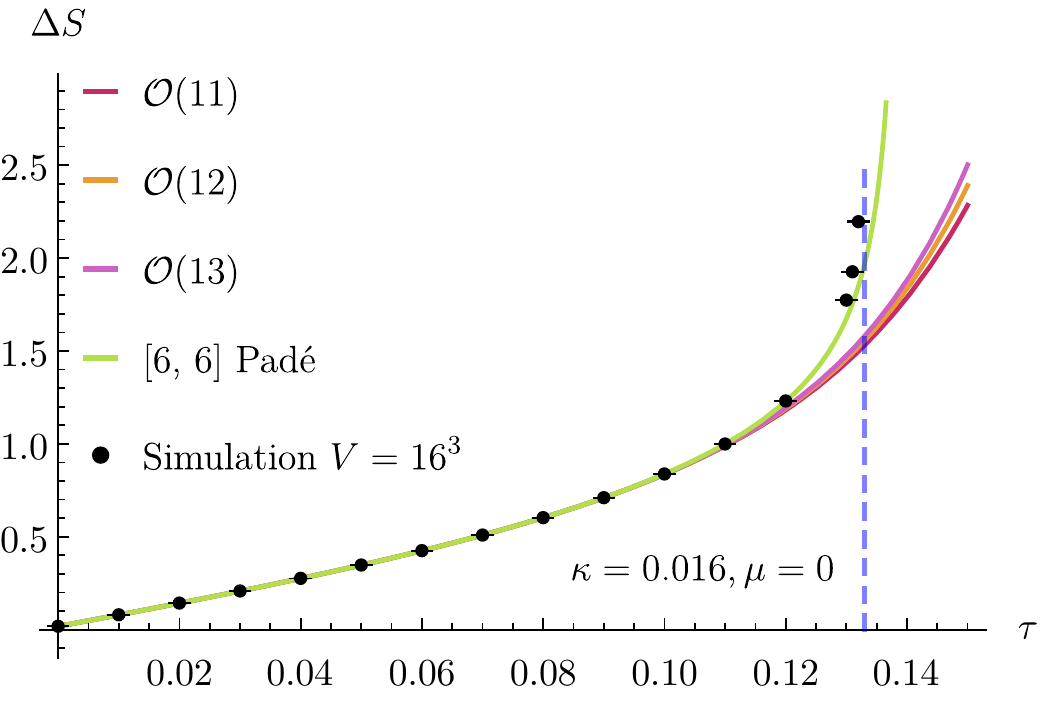}
 \includegraphics[width=0.5\textwidth]{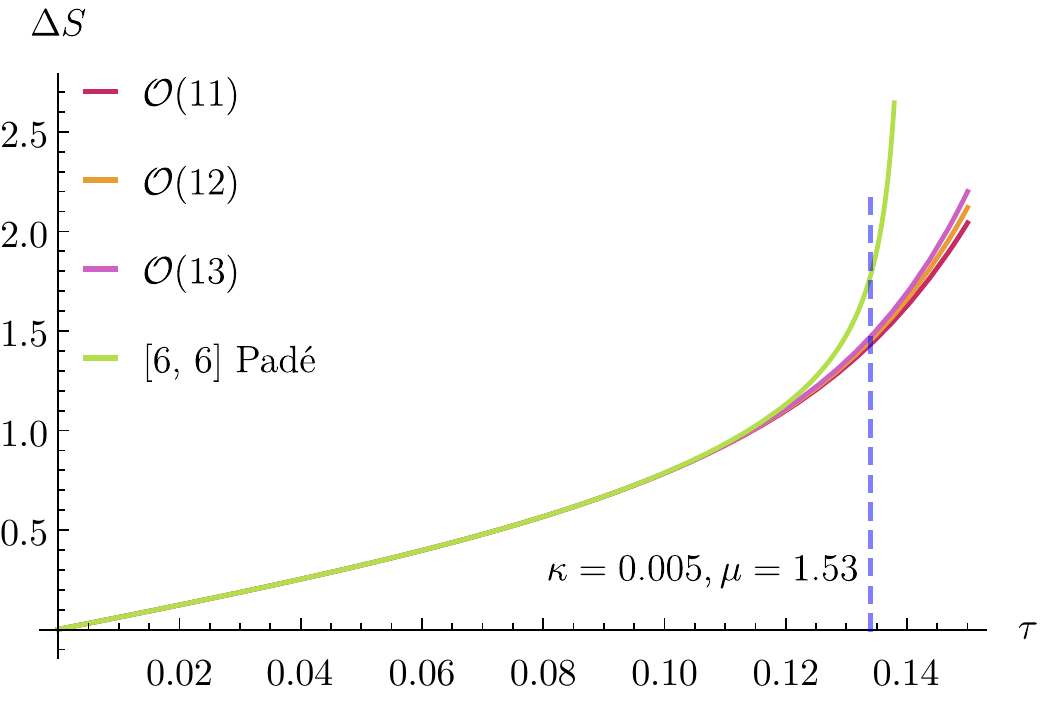}
  }
  \caption{The interaction measure from the highest three orders of our expansion. Left: At zero chemical potential, 
    compared to Monte Carlo data.
    Right: At large chemical potential, where standard Monte Carlo suffers from the sign problem.
     The critical $\tau$ is marked by the blue dashed line and based on simulations of a dual flux representation \cite{gat2}.}
  \label{fig:delta-s}
\end{figure}

The equation of state is often given in terms of the so-called interaction measure \eq \eqref{eq:int},
since the latter can be straightforwardly determined in Monte Carlo simulations. All thermodynamic
functions can be computed from it by integration or differentiation.
Our series results to the three highest orders 
are shown in \fig \ref{fig:delta-s}, where good convergence is observed.
At vanishing chemical potential (left), we 
compare our series directly to Monte Carlo data. We find excellent quantitative agreement 
up to the immediate neighbourhood of the phase transition, which a finite polynomial can
only indicate by loss of convergence. 
On the right we show results for a large chemical
potential, which poses no fundamental problem for a series expansion method, whereas standard
Monte Carlo simulations are no longer possible because of the sign problem. 

\subsection{Resummation by Pad\'e approximants}

Finite series generally break down in the vicinity of phase transitions. A marked improvement in convergence properties
can often be obtained by infinite-order resummations, like mappings of the expansion variables, use of renormalisation group
techniques or approximation by rational functions. For a general review and introduction, see \cite{Guttmann1989wz}.
Here we model a function $f(x)$, known only as finite power series, 
\beq
f(x)=\sum_{n=0}^N c_n x^n+{\cal O}(x^{N+1})\;,
\eeq
by Pad\'e approximants defined as rational functions,
\beq
[L,M](x)\equiv\frac{a_0+a_1x+\ldots+a_L x^L}{1+b_1x+\ldots+b_M x^M}\;.
\eeq
The coefficients $a_i, b_i$ are uniquely determined for $L+M \leq N$, if $N$ represents the highest available order 
of the expansion. In this way the
$[L,M]$ approximant reproduces the known series up to and including $\mathcal{O}(x^{L+M})$, where larger approximants
represent more expansion coefficients than smaller ones. As rational functions, Pad\'e approximants are able to show
singular behaviour and scaling properties near phase transitions. Quite generally, diagonal approximants with $L=M$ are expected to show
the best convergence properties, since they are invariant under Euler transformations of the expansion variable \cite{Guttmann1989wz},
i.e., the full function does not change under resummations of the power series caused by such transformations. 

In \fig \ref{fig:delta-s} we show the $[6,6]$ approximant to the ${\cal O}(\tau^{12})$ series of $\Delta S$. The phase transition
is now announced by a singularity in the approximant, and the equation of state is quantitatively accurate up to the transition.

\section{The phase transition  \label{sec:pda}}

In the previous section we have seen that the linked cluster expansion gives excellent results for the equation of state in 
the symmetric phase of  the $SU(3)$ spin model at zero as well as finite chemical potential. 
In this section we explore various methods to extract information about the location and order of the phase transition.  
Our preferred observable to locate phase transitions is the susceptibility, \eq \eqref{eq:susc}, which we know 
in the form of a power series through $\mathcal{O}(\tau^{14})$ and $\mathcal{O}(\kappa^{58})$, 
\beq
 \chi (\tau,\eta,\bar{\eta})= \sum_{n=0}^{14} c_n(\eta,\bar{\eta}) \,\tau^n\quad \Leftrightarrow \quad
  \chi (\tau,\kappa,\mu)= \sum_{n=0}^{14} c_n(\kappa,\mu) \,\tau^n\;.
 \label{eq:susc_series}
 \eeq
 Because of its definition through $\tau$-derivatives, the series for the specific heat, \eq \eqref{eq:specheat}, is by two orders
 shorter and, moreover, not related to the order parameter. We observe it to generally exhibit poorer convergence near the transition. 

\subsection{Radius of convergence from the ratio test and Pad\'e approximants}

\begin{figure}[t]
  \centerline{
    \includegraphics[width=0.5\textwidth]{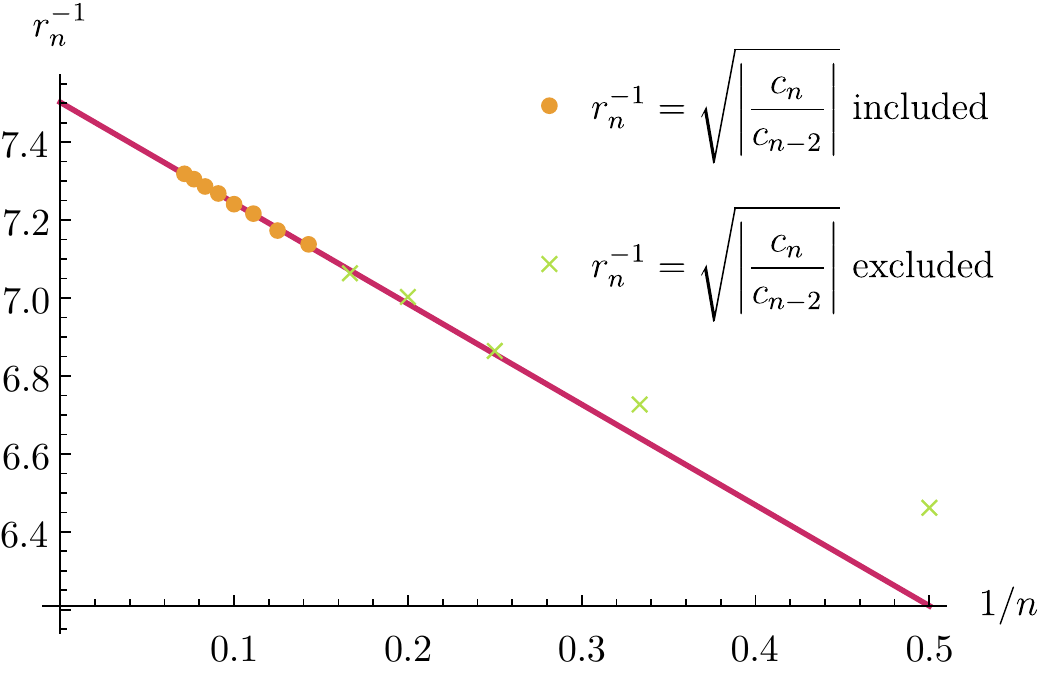}
    \includegraphics[width=0.5\textwidth]{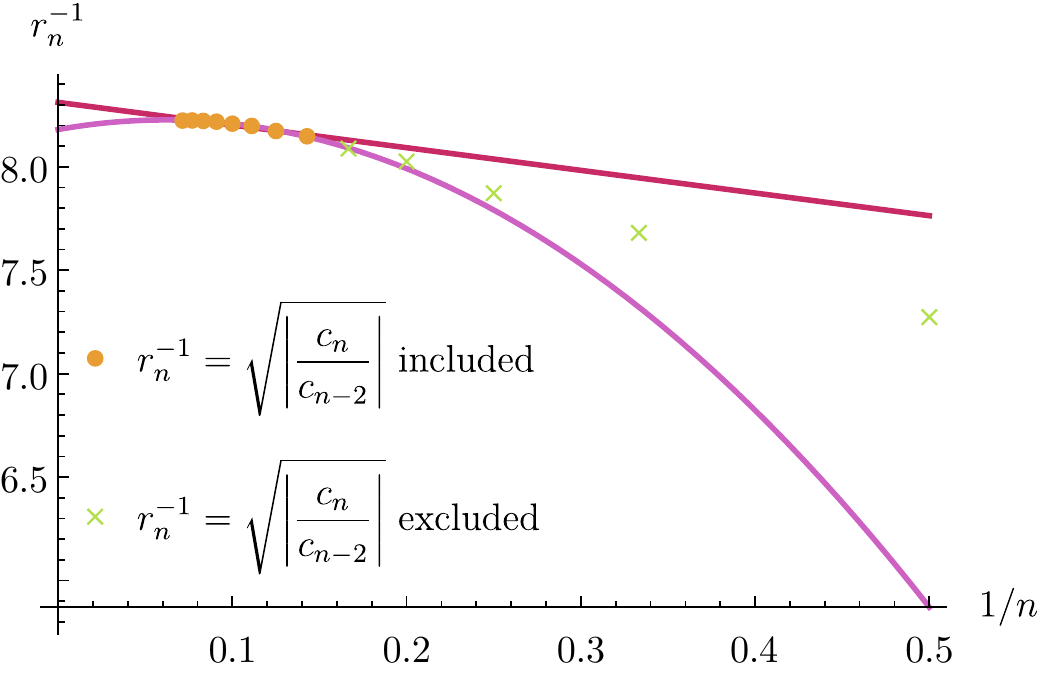}
  }
  \caption{Ratio test applied to the series \eqref{eq:susc_series} at $\mu=0$ for $\kappa=0.016$ (left) and $\kappa=0.08$ (right).
  Orange points are included in the fit.}
  \label{fig:ratio}
\end{figure}
When a function with a given domain of analyticity in its complex argument is expanded in a power series, the radius of convergence
is given by the distance between the expansion point and the nearest singularity. If this singularity is on the real positive
axis, it signals a phase transition. There are various estimators for the radius of convergence, which are appropriate for different 
circumstances. The simplest and most well-known is the ratio test of consecutive coefficients of a series, which
is expected to converge whenever coefficients have either the same or alternating signs. The radius of convergence is then obtained by extrapolation,
\beq
r=\lim_{n\to \infty} r_n\;,\quad r_n=\Big | \frac{c_n}{c_{n+1}}\Big|\;.
\eeq
Specifically, when there is a singularity on the positive real axis all
coefficients are positive for sufficiently large $n$, and if there are no
competing singularities then for large $n$ one expects the scaling
\cite{Guttmann1989wz}
\begin{equation}
  r_n^{-1} = 
  \frac{1}{r} 
  \left[
    1 + \frac{\lambda}{n} + \mathcal{O}\left(\frac{1}{n^2}\right)
  \right],
\end{equation}
which suggests a linear extrapolation to determine the radius of convergence.
(For sufficiently long series and second-order transitions, $\lambda$ is related to 
a critical exponent. Here we merely treat it as a fit parameter.)
We indeed observe a linear trend for the higher ratios for many points in the
parameter space, but with a superimposed oscillatory behaviour. This
is familiar  from Ising models on loose-packed lattices (in that case
due to an antiferromagnetic singularity) and can be reduced by using
the shifted ratios \cite{Gaunt:1974aa}
\begin{equation}
  r_n^{-1} = 
  \sqrt{\left| \frac{c_n}{c_{n-2}} \right|} =
  \frac{1}{r} 
  \left[
    1 + \frac{\lambda}{n} + \mathcal{O}\left(\frac{1}{n^2}\right)
  \right]\;.
\end{equation}
\Fig \ref{fig:ratio} shows the shifted estimators $r_n^{-1}$ plotted vs.~$n^{-1}$ for two values of $\kappa$ and $\mu=0$. 
As is apparent in the figure, a rather accurate estimate for the radius of convergence is obtained on the left, while
on the right the linear scaling region has not yet been reached.

\begin{figure}[t]
  \centerline{
    \includegraphics[width=0.5\textwidth]{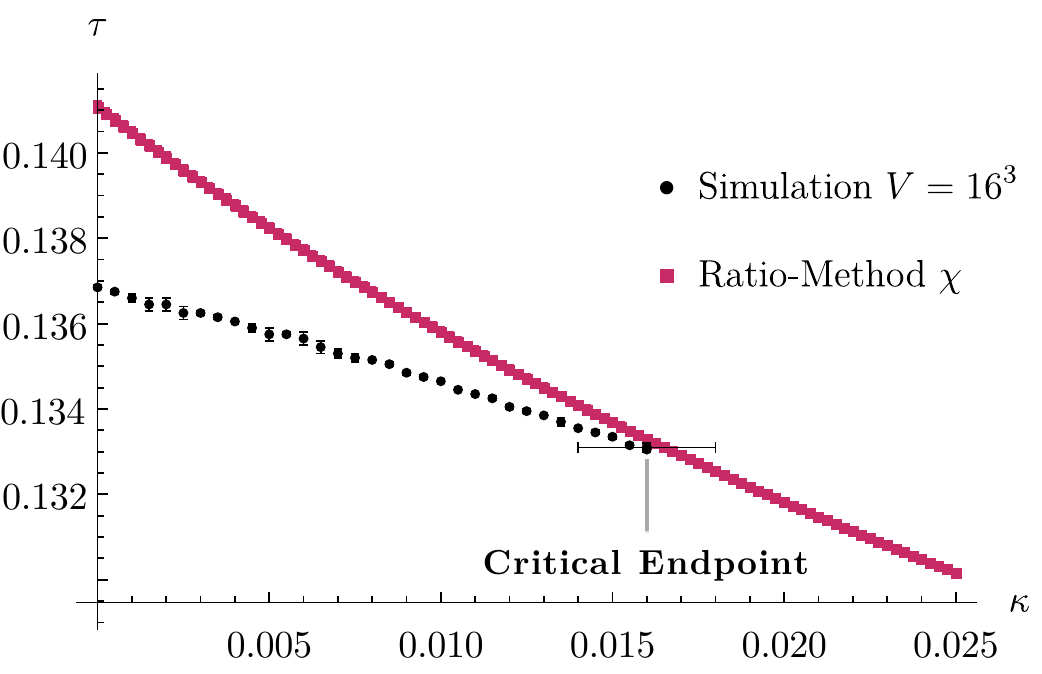}
    \includegraphics[width=0.5\textwidth]{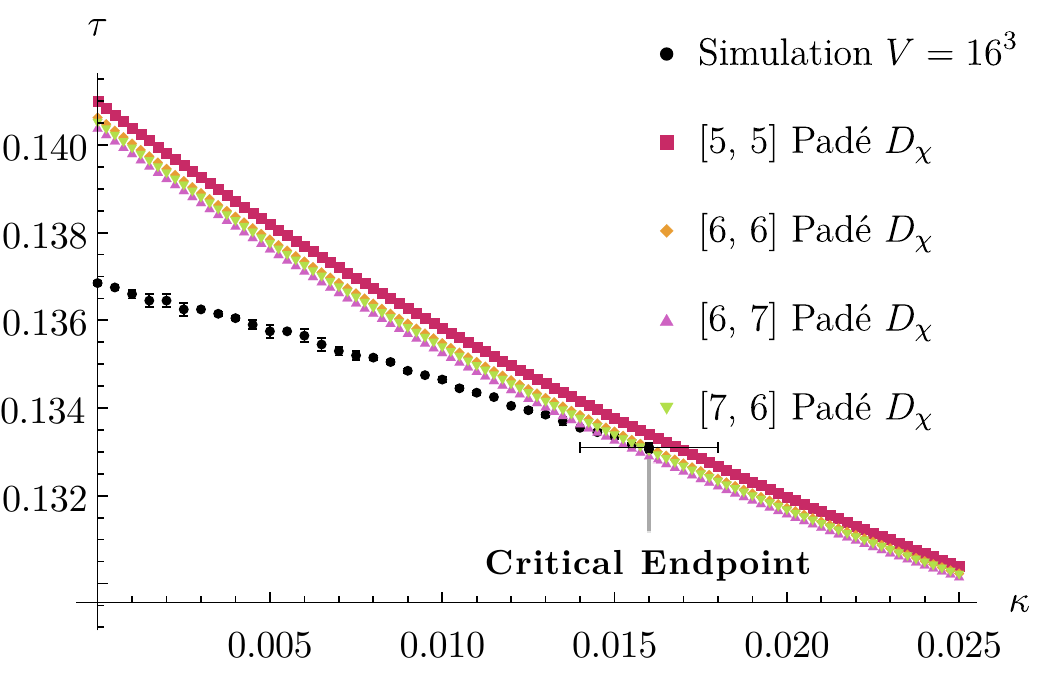}
  }
  \caption{Radius of convergence at $\mu=0$ from the ratio test (left) and Pad\'e approximants (right), 
  compared to the numerically determined phase diagram, with a first-order transition ending in a critical point. 
  In the first-order region, both approaches pick up the end of the 
  metastability region, and both are unable to distinguish crossover behaviour from a weak first-order transition.}
  \label{fig:crit_line}
\end{figure}

Repeating this procedure for varying values of $\kappa$, we obtain a line of singularities which is shown in \fig \ref{fig:crit_line} (left),
together with the phase structure obtained by Monte Carlo simulations. The latter show a line of first order transitions, which
weaken with $\kappa$ to end in a critical point, which has been computed in simulations of  
a flux representation \cite{gat2}. We observe the radius of convergence estimate from our series to systematically overshoot the critical coupling
by an amount which diminishes towards the critical endpoint, where it vanishes. This behaviour is due to the fact that 
the series has information from one phase only. Thus the singularity detected by a series analysis is the end of the metastability
region of that phase, rather than the true critical coupling, which requires information from both phases. The same behaviour is
observed in, e.g., Potts models \cite{Hellmund:2003vw} with first-order transitions.

A series analysis well-suited for second-order transitions is provided by Pad\'e approximants. At a second-order transition, the
susceptibility diverges with a critical exponent and, approaching the transition, its logarithmic derivative has a simple pole with 
the critical exponent as its residue,
 \beq
  \chi\sim \frac{1}{(\tau- \tau_c)^\epsilon}\;,\quad \Rightarrow \quad D_\chi(\tau)\equiv \frac{d}{d\tau}\log \chi\sim -\frac{\epsilon}{(\tau-\tau_c)}\;.
\eeq
Such a pole in the Dlog can be faithfully reproduced by Pad\'e approximants.  A pole in the full function $D_\chi$ is thus predicted whenever 
different approximants show converging roots of their denominators.

While Pad\'e approximants work astonishingly well for one-variable problems with a second-order transition (as a recent example, see
the strong coupling expansion of $SU(2)$ Yang-Mills theory \cite{Langelage:2009jb,Kim:2019ykj}), the two-variable situation of
the $SU(3)$ spin model with different types of transitions is more complicated. 
\Fig \ref{fig:crit_line} (right) shows the results
of these estimates, which are in remarkable agreement with those of the ratio test. 
The critical point is accurately reproduced at the appropriate $\kappa$-value, and the approximants also appear to pick
up the end of the metastability range in the first-order region. However, both the ratio test and the Pad\'e approximants indicate 
singularities also in the crossover region, where the full $\chi$ is known to be analytic.  
(With sufficiently long series, the radius of convergence should either become infinite or move to a complex value in this regime.)
In summary, both analyses  
pick up the rise of the susceptibility near the phase boundary, but are unable to clearly distinguish between orders of the phase transition
or crossover behaviour. Unfortunately, this difficulty remains when multi-variable generalisations of
Pad\'e approximants,
so-called partial differential approximants (see e.g.~\cite{Guttmann1989wz}), are used in both variables, 
as we have explicitly tested.

\subsection{The critical point}

\begin{figure}[t]
  \centerline{
    \includegraphics[width=0.45\textwidth]{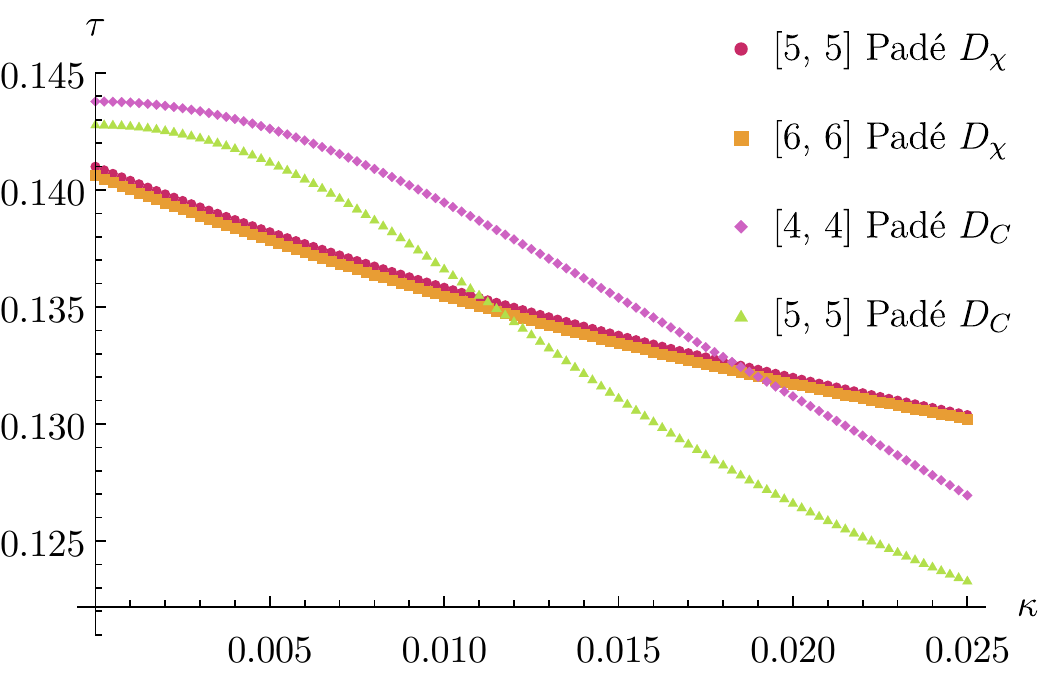} 
  }
  \caption{Location of the poles in the Pad\'e approximants to $D_C, D_\chi$ at $\mu=0$.}
  \label{fig:intersect}
\end{figure}
%
%
\begin{table}[hbt]
\centering
\begin{tabular}{|Sc|Sc|Sc|Sc|Sc|Sc|Sc|}
\hline
\multirow{9}*{$\mu=0.000$} & 
{Pad\'e $D_C$} & {Pad\'e $D_\chi$} & {$\tau_c$} & {$\kappa_c$} & {$\epsilon(C)$} & {$\epsilon(\chi)$}\\ 
\hline
&$[3,3]$ & $[4,4]$ & 0.13388& 0.01646& 0.67061& 0.71288\\ 
&$[3,3]$ & $[5,5]$ & 0.13322& 0.01681& 0.66652& 0.69369\\
&$[3,3]$ & $[6,6]$ & 0.13298& 0.01735& 0.66031& 0.66458\\
&$[4,4]$ & $[4,4]$ & 0.13365& 0.01639& 0.67143& 0.71287\\ 
&$[4,4]$ & $[5,5]$ & 0.13320& 0.01693& 0.66513& 0.69381\\
&$[4,4]$ & $[6,6]$ & 0.13268& 0.01733& 0.66053& 0.66453\\
&$[5,5]$ & $[4,4]$ & 0.13599& 0.01061& 0.62968& 0.71466\\
&$[5,5]$ & $[5,5]$ & 0.13546& 0.01107& 0.62009& 0.69108\\
&$[5,5]$ & $[6,6]$ & 0.13489& 0.01162& 0.60900& 0.65699\\
\hline
\hline
\multirow{7}*{$\kappa=0.005$} & 
{Pad\'e $D_C$} & {Pad\'e $D_\chi$} & {$\tau_c$} & {$\mu_c$} & {$\epsilon(C)$} & {$\epsilon(\chi)$}\\ 
\hline
&$[3,3]$ & $[4,4]$ & 0.13435& 1.82166& 0.64497& 0.72503\\ 
&$[3,3]$ & $[5,5]$ & 0.13159& 1.94362& 0.59861& 0.53562\\ 
&$[3,3]$ & $[6,6]$ & 0.12183& 2.23650& 0.47420& 0.07502\\ 
&$[4,4]$ & $[6,6]$ & 0.13848& 1.58232& 0.51950& 0.71739\\
&$[5,5]$ & $[4,4]$ & 0.13654& 1.32476& 0.60727& 0.72325\\ 
&$[5,5]$ & $[5,5]$ & 0.13604& 1.36700& 0.59123& 0.69755\\ 
&$[5,5]$ & $[6,6]$ & 0.12622& 1.83003& 0.39954& 0.07866\\ 
\hline
\end{tabular}
\caption[]{Intersection points of the poles in the Pad\'e approximants to $D_C, D_\chi$, together with the residues at the intersection.}
\label{tab:cross}
\end{table}

In order to locate the critical point, 
we now exploit the following facts: first, Dlog Pad\'es model the singular behaviour of the full observables correctly {\it only} at
a second order point; second, at such a critical point the susceptibility $\chi$ and the specific heat
$C$ show the same diverging behaviour. In fact, for $\eta, \bar{\eta}\neq 0$ the center symmetry of the spin model is explicitly broken, 
and the
$(\tau,\eta)$-axes are misaligned with the temperature and magnetic field scaling axes of the effective Ising Hamiltonian
governing the vicinity of the critical point.
This situation of a first-order transition terminating in a critical point is the generic one of a liquid-gas transition.
Consequently, if the critical point is approached in any direction asymptotically {\it not} parallel
to the first-order transition line, both quantities will diverge with the same mixed exponent  \cite{Griffiths:1970zz,Rehr:1973zz},
\beq
\chi\sim C\sim |\tau-\tau_c|^{-\epsilon}\;,\quad \epsilon=1-\frac{1}{\delta}=\frac{\gamma}{\beta\delta}\;.
\eeq
 For the 3d-Ising universality class,
 we have $\gamma=1.24, \beta=0.326, \delta=4.79$ and $\epsilon=0.79$.
  
\Fig \ref{fig:intersect} shows the location of the poles of the two highest-order approximants to each $D_\chi$ and $D_C$.
According to the arguments given above, these need to agree at a second-order transition and only there, 
so we take their intersections as
 estimates for the location of the critical point. 
 Note the comparatively worse convergence of the poles in $C$, which causes the largest part of the uncertainty.

We now systematise this analysis to obtain an error estimate.
 In \tbl \ref{tab:cross}  
 we list all intersections of diagonal Pad\'e approximants for $D_C$ and $D_\chi$ as estimates for the critical point. The last two 
 columns give the residue associated with each observable at the intersection point. 
 Pad\'e approximants necessarily produce ever more poles, the higher their order. It is then clear that some of them are
 artefacts and do not have physical meaning. To get rid of these, 
 we demand that the residues of the intersecting observables agree within 20\%.
 These estimates are then averaged over. 

\begin{figure}[t]
  \centerline{
    \includegraphics[width=0.45\textwidth]{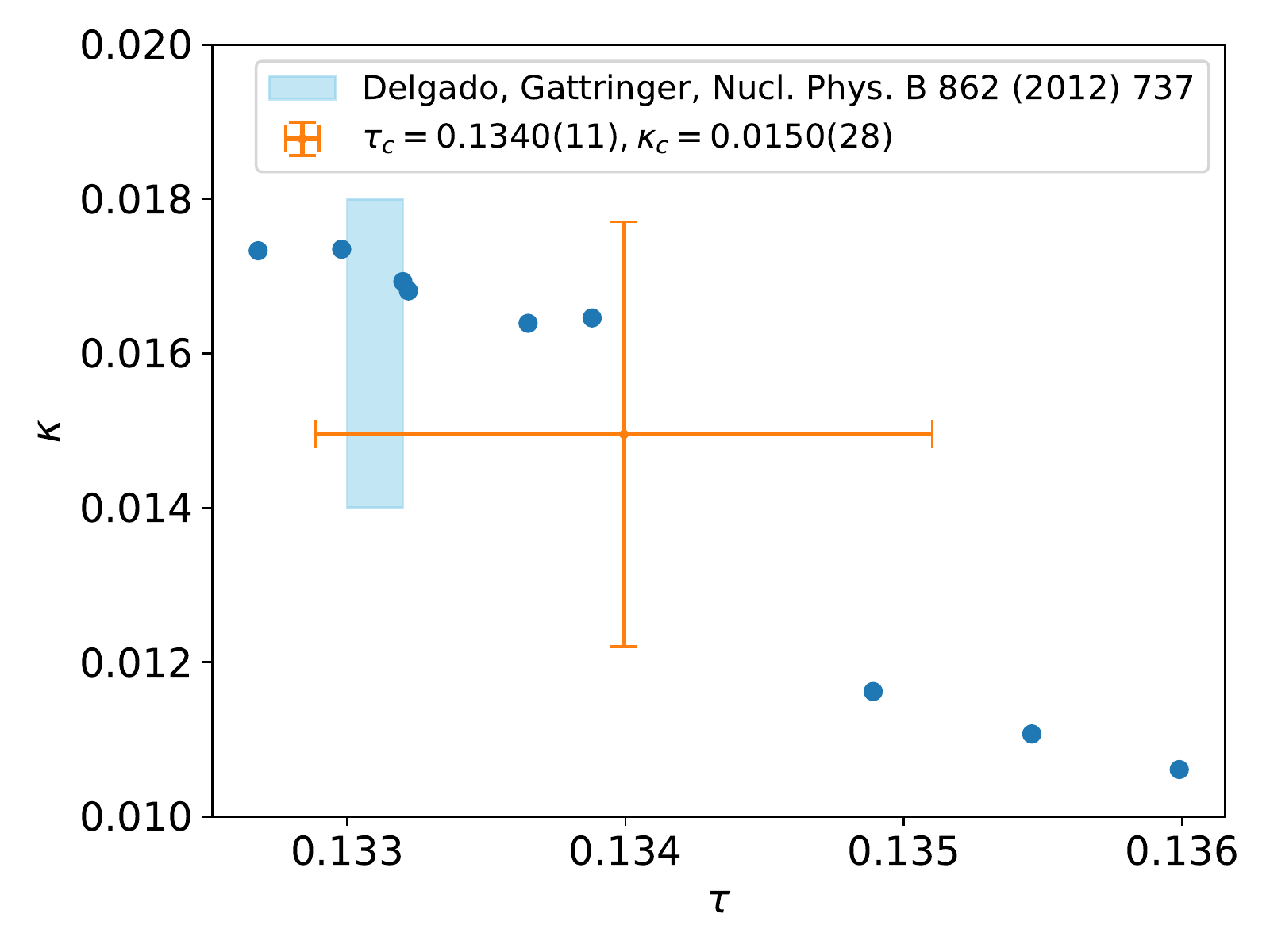} \hspace*{0.05\textwidth}
    \includegraphics[width=0.45\textwidth]{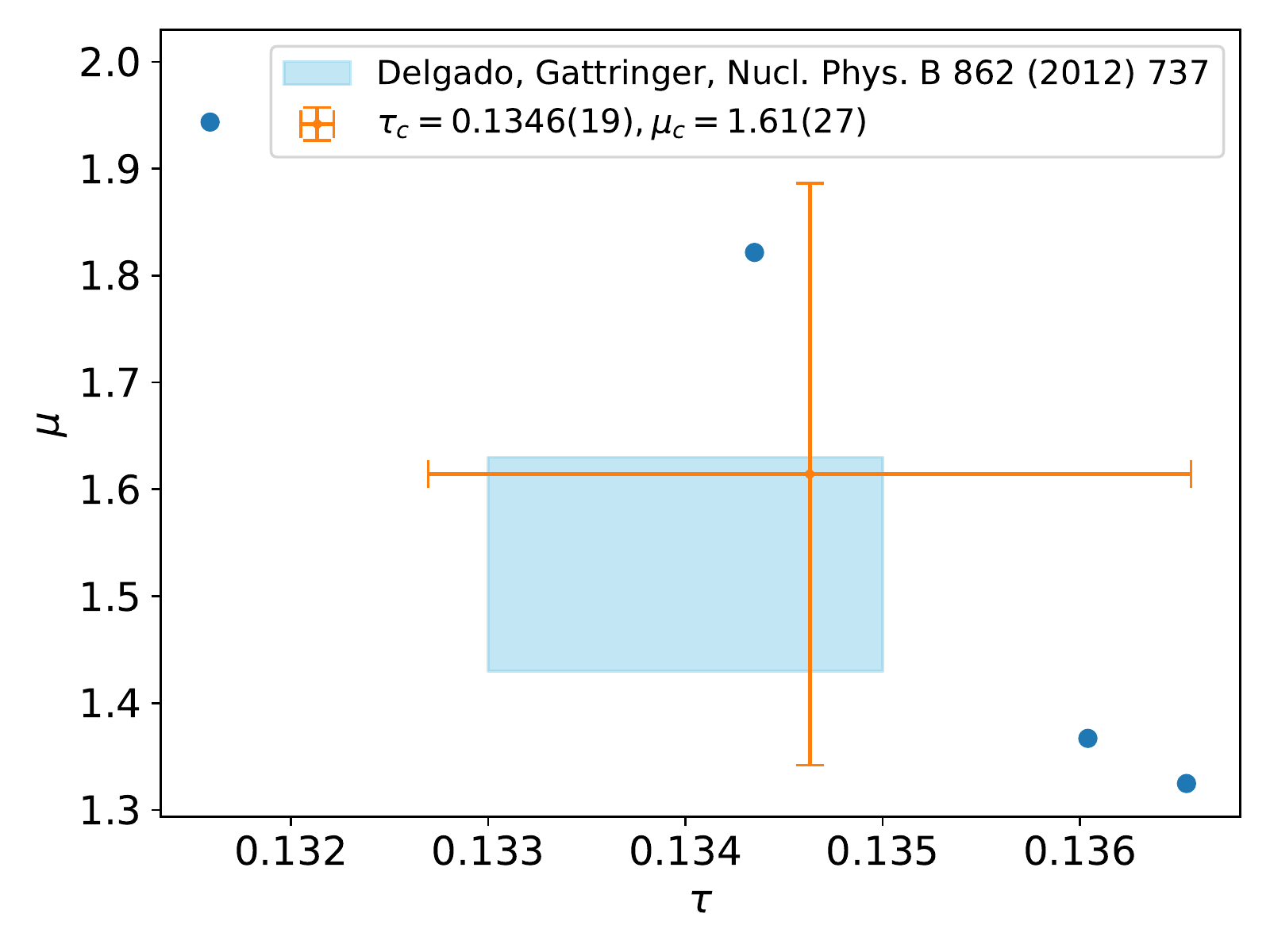}
  }
  \caption[]{Final estimates for the location of the critical point.  Left: $(\tau_c,\kappa_c)$ at $\mu=0$. Right: $(\tau_c,\mu_c)$ at 
  $\kappa=0.005$. Also shown are error bands from the numerical simulations in \cite{gat2}.}
  \label{fig:final}
\end{figure}

 The upper part of \tbl \ref{tab:cross} corresponds to
 $\mu=0$ as in \figs \ref{fig:crit_line}, \ref{fig:intersect}. Averaging over the different estimates gives our final result for the
 location of the critical point, which is shown in \fig \ref{fig:final} (left), in comparison to the simulation result from \cite{gat2}.
 The same kind of analysis can be repeated for finite chemical potential without further difficulties. 
 For the sake of comparison with the literature we now fix $\kappa$ and search for the critical point in $\mu$.
 The corresponding intersections of the pole positions in $D_C, D_\chi$ are collected in the lower part of \tbl \ref{tab:cross}.
 Upon inspection, one identifies the $[4,4]-[6,6]$ and $[5,5]-[6,6]$ crossings as spurious, because the associated
 residues deviate drastically from each other and the other values in the table. This leaves us with four estimates to be
 averaged, as shown in \fig \ref{fig:final} (right), again in comparison to the numerical result from \cite{gat2}. 
 The predictions from the series approach have a relative error of about 10\% in $\tau_c$ and about 20\% in $\kappa_c$ or $\mu_c$.
 Presumably the larger error on the latter variables is due to the flatness of the critical line in the phase diagram, 
 \fig \ref{fig:crit_line}, so it takes more accuracy (and thus higher orders) to resolve changes in those directions.
Within error bars, the predicted critical points agree with the numerically determined ones, so the estimate of the systematic 
error is realistic. Finally, we compute the critical exponent from the more reliable $D_\chi$ and find
$\epsilon=0.69(2)$ and $\epsilon=0.67(8)$ for the two computed critical points, respectively. This is within $\sim$15\% of the true 
value. 

\subsection{Real and imaginary chemical potential}

It is easy to see from equation \eqref{eq:su3spinaction} that for purely imaginary chemical potential,
$\mu=i\mu_i,\mu_i\in \mathbb{R}$, the action of the $SU(3)$ spin model is real and thus has no sign problem. 
The partition function is then periodic in imaginary chemical potential and exhibits the Roberge-Weiss symmetry 
shared by QCD \cite{Roberge:1986mm},
\beq
Z(\tau,\kappa,i\mu_i)=Z(\tau,\kappa,i\mu_i+i2\pi n/3)\;, \quad n=0,1,2,\ldots\;,
\eeq
i.e., center transformations are equivalent to shifts in imaginary chemical potential. Moreover, the partition function
is an even function of $\mu$, $Z(\mu)=Z(-\mu)$, so that the free energy density or the pressure can be expanded in powers of $\mu^2$.
Analytic continuation between real and imaginary $\mu$ is then trivial and can be utilised for sufficiently small chemical 
potential. This provides another test for our computational method, which in principle does not distinguish between real
and imaginary chemical potential. Of course, the convergence properties of the series are affected by the choice of parameter
values and may well be different in different directions of parameter space.

\begin{figure}[t]
  \centerline{
    \includegraphics[width=0.45\textwidth]{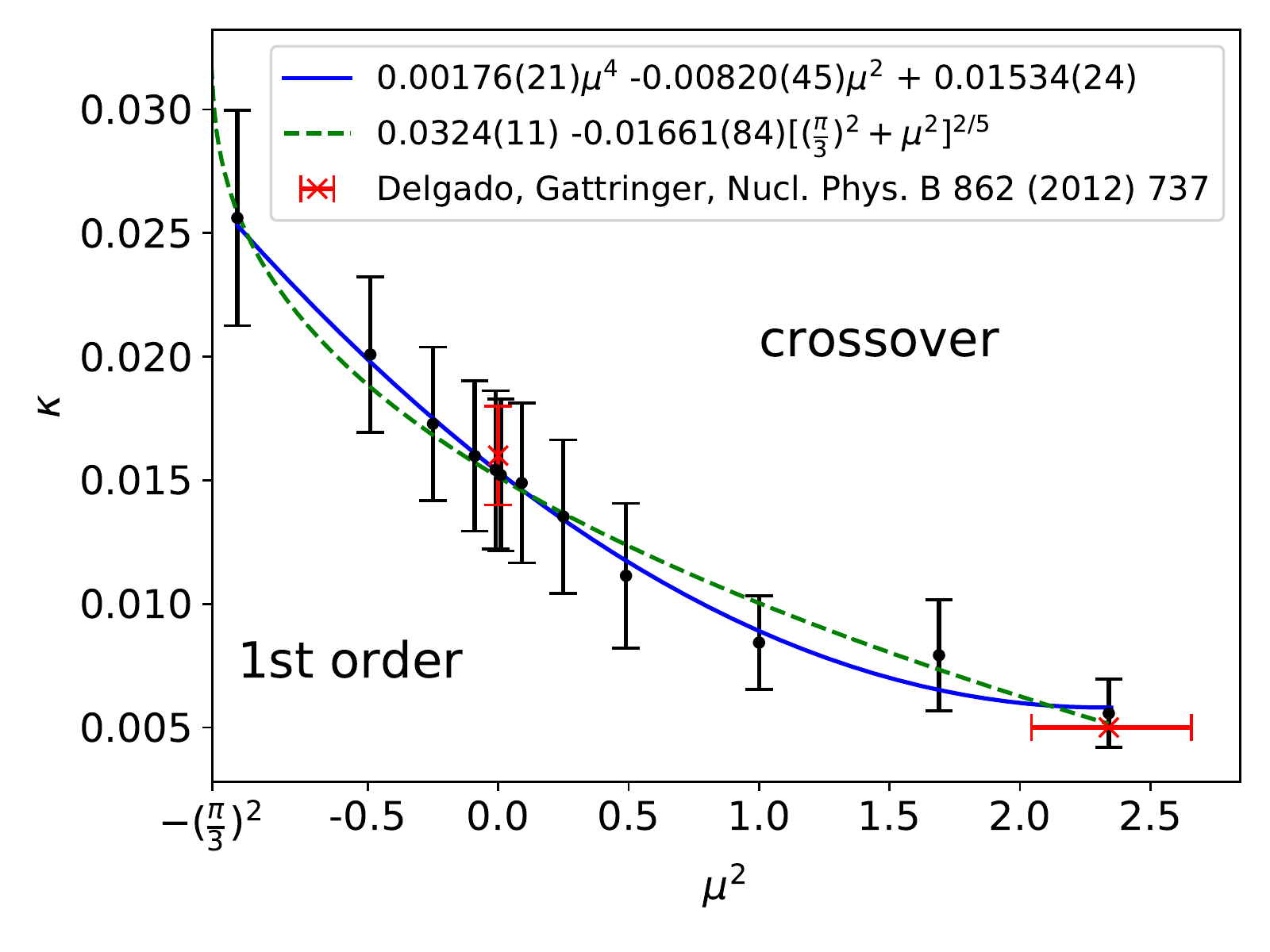}
  }
  \caption[]{Order of the center symmetry transition as a function of $(\kappa,\mu)$. A critical line of endpoints separates the first-order 
  region from the crossover region of parameter space, cf.~figure \ref{fig:schem}. Red data points are from numerical simulations,
  lines correspond to fits as described in the text. }
  \label{fig:immu}
\end{figure}

We have repeated our analysis described in the last section for a series of parameter values and mapped out how the location of the critical
endpoint changes as a function of $\kappa$ and $\mu$. Thus the complete phase diagram of the theory is determined,
as shown in figure \ref{fig:immu}. For every choice of $(\kappa,\mu)$ there is a $\tau_c(\kappa,\mu)$ marking the phase 
boundary for center symmetry transition. The figure shows the second-order critical line separating the parameter region with 
first-order transitions from that of smooth crossovers, i.e., it is a projection into the transition surface of figure \ref{fig:schem}.  
The expected 
analyticity of the critical line around $\mu=0$ is clearly observed, and at this moderate accuracy the entire range of chemical potentials is well 
described by fitting a next-to-leading-order Taylor expansion in $\mu^2$ about zero. 
Note that at $-\mu^2=(\pi/3)^2=(1.05)^2$ the boundary to the neighbouring center sector
is crossed, beyond which the phase diagram is dictated by the Roberge-Weiss symmetry. 
This point is marked by a cusp, where two critical lines from neighbouring center sectors meet, and which thus is tricritical
in all theories featuring this center symmetry, such as the $Z(3)$ Potts model or QCD \cite{deForcrand:2010he}.
As a consequence, the critical line is leaving this point with a known tricritical exponent, and a fit to this functional
form also describes our results over the whole range.
 
Finally, we remark that the parameter range 
$0.016<\kappa < 0.03$ corresponds to a situation, where there is a crossover for $\mu=0$ but a critical point followed by a first-order
transition beyond some imaginary $\mu_c$, as is often expected for QCD and real $\mu$. 
We conclude that the computational technique and analysis examined 
here is in principle able to handle such a situation.

\section{Conclusions \label{sec:conc}}

We have studied the $SU(3)$ spin model with chemical potential employing analytic series expansion techniques. 
In particular, we determined 
the free energy density through
order $\tau^{14}$ in the coupling of the energy-like term and through order $\kappa^{60}$ in the magnetisation-like coupling. In the phase 
with unbroken center symmetry, where the expansion is valid, we observe excellent convergence and the numerically known 
equation of state is reproduced
to  $\sim$1\% up to the phase transition. The series holds for both zero and 
finite chemical potential, since it is unaffected by sign problems.

We then investigated different approaches to extract information about the phase transition from the power series.
Agreeing estimates for the radius of convergence were obtained by extrapolations of the ratio test and Pad\'e approximants 
to the series of the susceptibility in
the $\tau$-variable. In the parameter region with a first-order transition, the detected singularity corresponds to
the end of the metastability region of the symmetric phase. At the critical end point of the transition, 
the estimates agree with the true transition,
but finite radius of convergence estimates continue far into the crossover region, where there is no singularity for real couplings.  
Such methods by themselves are therefore unable to determine the nature of a phase transition in a multi-variable system.
On the other hand, exploiting the fact that the magnetic susceptibility and the specific heat diverge in the same manner 
at the critical point, an analysis of the crossings of their respective Dlog Pad\'es allows to extract estimates for the location of the
critical point with a relative accuracy of 10-20\% in the critical couplings. In this manner the entire phase diagram for zero,
imaginary and real chemical potential could be mapped out successfully. 

In comparison to numerical solutions via flux representations, the current depth series
is competitive for the equation of state, but less accurate for the critical couplings. 
Never\-theless, it adds flexibility and offers an interesting perspective to finite density QCD, where such simulations are not yet applicable.
In particular, our results show that effective theory approaches based on strong coupling and hopping
 expansions \cite{llp,fromm,bind,Glesaaen:2015vtp}
are reliable and, with some effort to develop efficient computational schemes for higher orders and multiple couplings, 
can be systematically improved to apply 
to larger QCD parameter regions.   

\acknowledgments We thank W.~Unger for discussions and the permission to use
the ARIADNE code suite.
We acknowledge support by the Deutsche Forschungsgemeinschaft (DFG) through the grant CRC-TR 211 ``Strong-interaction matter
under extreme conditions''.

\appendix
\section{Computational steps and run-times}
\label{sec:a}

The computation of the free energy density in the renormalised scheme consists of the following steps:
\begin{enumerate}
  \item Enumeration of all relevant graphs, computation of their
    symmetry embedding numbers. We only generate undirected graphs
    in this step, the different directed graphs are taken into account
    when the graphs are converted to expressions containing the
    self-fields and semi-invariants.
  \item Computation of the renormalised semi-invariants in terms of
    the self-fields and bare semi-invariants.
  \item Determination of the self-fields in terms of the renormalised semi-invariants.
  \item Evaluation of the bare semi-invariants in terms of $\eta$ and $\bar{\eta}$ up to order
    $\mathcal{O}(\kappa^{60})$.
  \item Determination of the self-fields and renormalised semi-invariants in terms
    of $\eta$ and $\bar{\eta}$ using the bare semi-invariants from the previous step
    and the iterative procedure explained in section \ref{subsec:vertex-renorm}. This step is 
    computationally the most expensive part, as one repeatedly has to expand products
    of series in $\eta$ and $\bar{\eta}$.
  \item Determination of the $\Phi$-functional in terms of the self-fields and renormalised
    semi-invariants.
  \item Determination of the free energy in terms of $\eta$ and $\bar{\eta}$.
\end{enumerate}
In its current form, the implementation is certainly far from optimal in terms
of performance, nevertheless we give the run-time of these steps to the
computed orders on a Intel(R) Core(TM) i5-7500 CPU @ 3.40GHz in \tbl \ref{tab:runningtime},
to indicate the general behaviour when increasing the order.

\begin{table}[htb]
  \begin{tabular}{c|rrrrrr}
    Step & $\mathcal{O}(\tau^{10})$ & $\mathcal{O}(\tau^{11})$ & $\mathcal{O}(\tau^{12})$ &
    $\mathcal{O}(\tau^{13})$ & $\mathcal{O}(\tau^{14})$ \\
    \hline
    1 & 7\text{s} & 14\text{s} & 41\text{s} & 1\text{m} 57\text{s} & 8\text{m} 38\text{s} \\
    2 & 5\text{s} & 10\text{s} & 23\text{s} & 52\text{s} & 1\text{m} 57\text{s} \\
    3 & 19\text{s} & 1\text{m} 8\text{s} & 5\text{m} 23\text{s} & 29\text{m} 25\text{s} & 
    3\text{h} 8\text{m} 26\text{s} \\
    4 & 11\text{m} 58\text{s} & 13\text{m} 59\text{s} & 16\text{m} 20\text{s} & 
    19\text{m} 0\text{s} & 22\text{m} 8\text{s} \\
    5 & 1\text{h} 5\text{m} 8\text{s} & 3\text{h} 26\text{m} 18\text{s} & 
    11\text{h} 24\text{m} 20\text{s} & 37\text{h} 48\text{m} 0\text{s} &
    124\text{h} 25\text{m} 26\text{s} \\
    6 & 6\text{s} & 20\text{s} & 1\text{m} 24\text{s} & 6\text{m} 23\text{s} &
    35\text{m} 7\text{s} \\
    7 & 27\text{m} 4\text{s} & 1\text{h} 33\text{m} 28\text{s} &
    5\text{h} 28\text{m} 44\text{s} & 18\text{h} 21\text{m} 56\text{s} & 
    59\text{h} 50\text{m} 6\text{s} \\
    \hline
    $\Sigma$ & 1\text{h} 44\text{m} 47\text{s} & 5\text{h} 15\text{m} 38\text{s} &
    17\text{h} 17\text{m} 16\text{s} & 57\text{h} 7\text{m} 33\text{s} & 
    188\text{h} 31\text{m} 48\text{s}
  \end{tabular}
  \caption{Run-time to compute the free energy to $\mathcal{\kappa}^{60}$ and different
  orders in $\tau$.}
  \label{tab:runningtime}
\end{table}

\end{document}